\documentclass[sigconf]{acmart}

\usepackage{listings}
\usepackage{xcolor}

\AtBeginDocument{%
  \providecommand\BibTeX{{%
    \normalfont B\kern-0.5em{\scshape i\kern-0.25em b}\kern-0.8em\TeX}}}

\setcopyright{acmlicensed}
\copyrightyear{2025}
\acmYear{2025}
\acmDOI{XXXXXXX.XXXXXXX}

\acmConference[Conference acronym 'XX]{Make sure to enter the correct
  conference title from your rights confirmation email}{June 03--05,
  2018}{Woodstock, NY}
\acmISBN{978-1-4503-XXXX-X/18/06}

\graphicspath{{./images/}}

\begin{document}

\title{JOINT – Join Optimization and Inference via Network Traversal}

\author{Szu-Yun Ko}
\email{b12705066@ntu.edu.tw}
\affiliation{%
  \institution{Department of Information Management}
  \city{National Taiwan University}
  \country{Taiwan}
}

\author{Bo-Cian Chang}
\email{b09508009@ntu.edu.tw}
\affiliation{%
  \institution{Department of Economics}
  \city{National Taiwan University}
  \country{Taiwan}
}

\author{Alan Shu-Luen Chang}
\email{b11501010@ntu.edu.tw}
\affiliation{%
  \institution{Department of Civil Engineering}
  \city{National Taiwan University}
  \country{Taiwan}
}

\author{Ethan Chen}
\email{r13921091@ntu.edu.tw}
\affiliation{%
 \institution{Department of Electrical Engineering}
 \city{National Taiwan University}
 \country{Taiwan}
}
\renewcommand{\shortauthors}{Ko, et al.}

\begin{abstract}
  Traditional relational databases require users to manually specify join keys and assume exact matches between column names and values. In practice, this limits joinability across fragmented or inconsistently named tables. We propose a fuzzy join framework that automatically identifies joinable column pairs and traverses indirect join paths across multiple databases. Our method combines column name similarity with row-level fuzzy value overlap, computes edge weights using negative log-transformed Jaccard scores, and performs join path discovery via graph traversal. Experiments on synthetic healthcare-style databases demonstrate the system’s ability to recover valid joins despite fuzzified column names and partial value mismatches. This research has direct applications in data integration.
\end{abstract}


\maketitle

\section{Introduction}
In traditional relational database systems, join operations require users to explicitly specify matching columns, typically relying on exact name and value equivalence. Even \texttt{NATURAL JOIN}, which infers join keys automatically, is limited to columns with identical names and compatible data types, and still assumes exact value matches. These rigid constraints often fall short in real-world scenarios, where datasets may exhibit inconsistent naming conventions, abbreviations, or contain typos and semantic variations in both schema and content.

To address these limitations, we aim to enable a more flexible and realistic join discovery framework that can handle fuzzy column names and record values. In addition, we intend to enable the system to discover indirect join paths through multiple tables. 

This capability is particularly valuable in applications such as data integration, exploratory data analysis, and automated schema mapping, where datasets from different sources often lack standardized structures.

By relaxing the reliance on exact matches, we move toward a more robust interpretation of table relationships in relational systems.

\section{Related Works}
\subsection{Fuzzy Row-Level Matching}
Fuzzy joins, or similarity joins, have been widely studied in the context of matching tuples with non-identical values. Traditional approaches typically require users to manually specify a join function and a similarity threshold, which can be complex and error-prone. Auto-FuzzyJoin~\cite{li2021auto} addresses this by automatically generating high-quality fuzzy join programs without labeled training data. It learns appropriate distance functions and thresholds for a given pair of columns, achieving high precision and recall. However, Auto-FuzzyJoin assumes that the matching columns are already known and focuses only on single-step joins between two tables. Also, Auto-FuzzyJoin heavily depends on the assumption that a reference table with nearly no duplicate values exist in the DBMS, which may not be realistic for all DBMS designs.

\subsection{Entity Matching with Language Models}

Recent work has explored using large language models (LLMs) for entity matching across datasets. Peeters et al.~\cite{peeters2025entity} evaluate the effectiveness of generative LLMs, such as GPT-4, Gemini and LLaMA, in both zero-shot and few-shot scenarios. They demonstrate that LLMs can perform competitively with traditional pre-trained models like RoBERTa, even without task-specific training data. However, these models operate at the tuple level and require prompts to be manually constructed. The system is also highly sensitive to prompt wording, and different prompts are required for different context and datasets, there is currently no general prompt that may suit all scenarios. Also, this system does not generalize to multi-hop joins or automatically infer which columns or paths to consider.

\subsection{Our Contribution}

As we identified some gaps in the current literature, we attempt to bridge these gaps with our JOINT system. Our system integrates fuzzy column name matching with row-level similarity comparisons to automatically identify joinable column pairs. Furthermore, we extend this capability to discover multi-table join paths using a weighted graph traversal approach. Our method relaxes the need for exact matches at both the column and row levels, supporting more robust join discovery in real-world, messy datasets.

\section{Research Objectives} 
\begin{enumerate}
    \item \textbf{Fuzzily matching column names} to identify semantically similar attributes across tables, even when their names differ (e.g., \texttt{`doctor\_name`} vs. \texttt{`assigned\_doctor\_name`}).
    \item \textbf{Identify fuzzy record matches} across databases despite discrepancies in notation (e.g., typos, abbreviations, different word ordering...etc.). 
    \item \textbf{Discovering indirect join paths} by traversing through intermediate tables based on fuzzy match results, allowing us to infer joinability even when no direct foreign key relationship exists.
\end{enumerate}

\section{Methods}

To achieve the research objectives, we propose a multi-stage methodology that integrates metadata reflection, column-level and row-level fuzzy similarity scoring, and graph-based join path construction. The overall process is composed of five main stages:

\subsection{Metadata Extraction and Identification of Foreign Key Edges}

\textbf{Purpose:} Establish the baseline join graph using foreign key (FK) constraints between tables within individual DBMS.\\
\textbf{Implementation:} 
We use \texttt{SQLAlchemy} to extract FK constraints for tables in the same DBMS. Foreign key relations between two tables will be converted into FK edges between the respective tables in the join graph.

\subsection{Fuzzy Column Matching Across DBMS}

\textbf{Purpose:} Detect column pairs with high semantic similarity despite inconsistent naming across databases (e.g., \texttt{`drug\_name`} vs \texttt{`medication\_name'}).\\
\textbf{Implementation:} Candidate column pairs are scored using a weighted combination of three metrics:
    \begin{enumerate}
        \item Name similarity using \texttt{SequenceMatcher}.
        \item Token overlap ratio, which is order-insensitive.
        \item Semantic similarity from transformer embeddings\\ (\texttt{all-MiniLM-L6-v2}) using cosine distance \cite{reimers-2021-minilm}.
    \end{enumerate}
Each score is normalized to [0, 1] and aggregated using:
    \[
        \text{total\_score} = \alpha \cdot \text{name\_sim} + \beta \cdot \text{semantic\_sim} + \gamma \cdot \text{token\_overlap}
    \]
where \(\alpha + \beta + \gamma = 1\). Only pairs exceeding a threshold (e.g., 0.6) are retained. All parameters and threshold can be defined by the user.

\subsection{Row-Level Fuzzy Validation}

\textbf{Purpose:} Confirm column pair joinability by measuring actual row-level value similarity.\\
\textbf{Implementation:} We compute fuzzy similarity scores between value lists using \texttt{rapidfuzz}'s \texttt{token\_sort\_ratio} \cite{rapidfuzz}. The row-wise similarity matrix is reduced by taking the maximum score per row and then averaged:
    \[
        \text{value\_score} = \frac{1}{n} \sum_{i=1}^n \max_j \text{sim}(x_i, y_j)
    \]
Pairs with average scores above a second threshold (e.g., 0.5) are appended to the join graph as fuzzy edges.

\subsection{Weight Assignment to FK and Fuzzy Edges in Join Graph}
Column overlap $s_i$ denotes the record overlap between the $i^{th}$ and $(i+1)^{th}$ table in the join sequence, $s_i$ is computed as: 
    $$s_{i}=\text{Jaccard}(C_i, C_{i+1}) = \frac{|C_i \cap C_{i+1}|}{|C_i \cup C_{i+1}|}$$ 
    We aim to maximize cumulative overlap across multiple tables, since overlap is a percentage measure, cumulative overlap is computed as a product of each pair-wise overlap in the join sequence:
    $$\max\prod_{i=1}^ns_i$$
    therefore we define the edge weight as: 
    $$\text{weight}_i = -\log(s_i+\epsilon)$$
The pathfinding algorithm will aim to minimize cumulative edge weights (shortest path), which is equivalent to maximizing cumulative overlap across multiple tables:
    $$\min\sum_{i=1}^nw_i=\min-\log(\prod_{i=1}^n(s_i+\epsilon))$$
A higher overlap implies stronger joinability and thus a lower traversal weight. Edges are added to a \texttt{networkx} graph with associated columns and scores.
This edge weight definition ensures additive cost under Dijkstra traversal.

\subsection{Join Graph Construction and Path Discovery}
\textbf{Purpose:} Form a heterogeneous join graph and identify optimal multi-hop join paths for analytics.\\
\textbf{Implementation:} Nodes in the graph represent tables across all DBMSs. Edges include FK joins and validated fuzzy joins. Each edge carries metadata: join type, columns, similarity, and traversal weight. We apply Dijkstra's algorithm to compute shortest paths between any two target tables. The effective retained data can be approximated as:
    \[
        \text{retained\_percentage} \approx 2^{-\text{total\_weight}}
    \]
This favors paths with stronger join overlaps and higher semantic integrity.

\section{Experimentation}
\subsection{Synthetic Database Design}

We construct a synthetic multi-database environment consisting of the following four datasets:
\begin{itemize}
    \item \textbf{hospital\_db} – Contain five tables: `Patients`, `Clinics`, `Prescriptions`, `Doctors` and `Appointments`.
    \item \textbf{insurance\_db} – Contain three tables: `Insurance\_Providers`, `Insured\_Patients` and `Claims`.
    \item \textbf{pharmacy\_db} – Contain three tables: `Pharmacies`, `Drugs`, and `Pharmacy\_Orders`.
    \item \textbf{public\_info\_db} – Contain three tables: `Citizen\_Registry`, `Hospital\_Survey`, and `Drug\_Watchlist`.
\end{itemize}
The generator functions, detailed tables and records of hospital\_db, insurance\_db and pharmacy\_db can be found in the attached code \texttt{`healthcare\_data.ipynb`}. Details of public\_info\_db can be found in \texttt{`add\_fuzzy\_db.ipynb`}.

Each database simulates independently maintained systems with partial schema overlap. We currently assume that all appropriate foreign key constraints are properly configured. 

\subsection{Fuzzification of Records}

To simulate real-world noise, we apply controlled fuzzification to a subset of the records. The transformations include:
\begin{enumerate}
    \item \textbf{Character removal:} Randomly delete 1–2 characters from a name string to simulate typos.
    \item \textbf{Name reordering:} Reverse name parts, e.g., ``John Smith'' becomes ``Smith John''.
    \item \textbf{Synonym injection:} Replace drug names using a predefined mapping (e.g., ``Amoxicillin'' $\rightarrow$ ``Amoksillin'').
    \item \textbf{Label variation:} Append suffixes like ``Clinic'' or ``Hospital'' to simulate entity naming inconsistency.
\end{enumerate}

These utilities are implemented as randomized functions (see \texttt{`add\_fuzzy\_db.ipynb`}) and applied selectively during data generation.

\subsection{Join Graph Construction}

We build a join graph where:
\begin{itemize}
    \item Each node represents a table.
    \item Edges represent joinable column pairs, scored by fuzzy matching (e.g., token overlap, embedding similarity).
    \item Edge weights are computed as mentioned in section 4.4.
\end{itemize}

The join graph built upon our synthetic data warehouse consisting of four DBMS is as shown in Figure 1:
\begin{figure}[ht]
    \centering
    \includegraphics[width=0.9\linewidth]{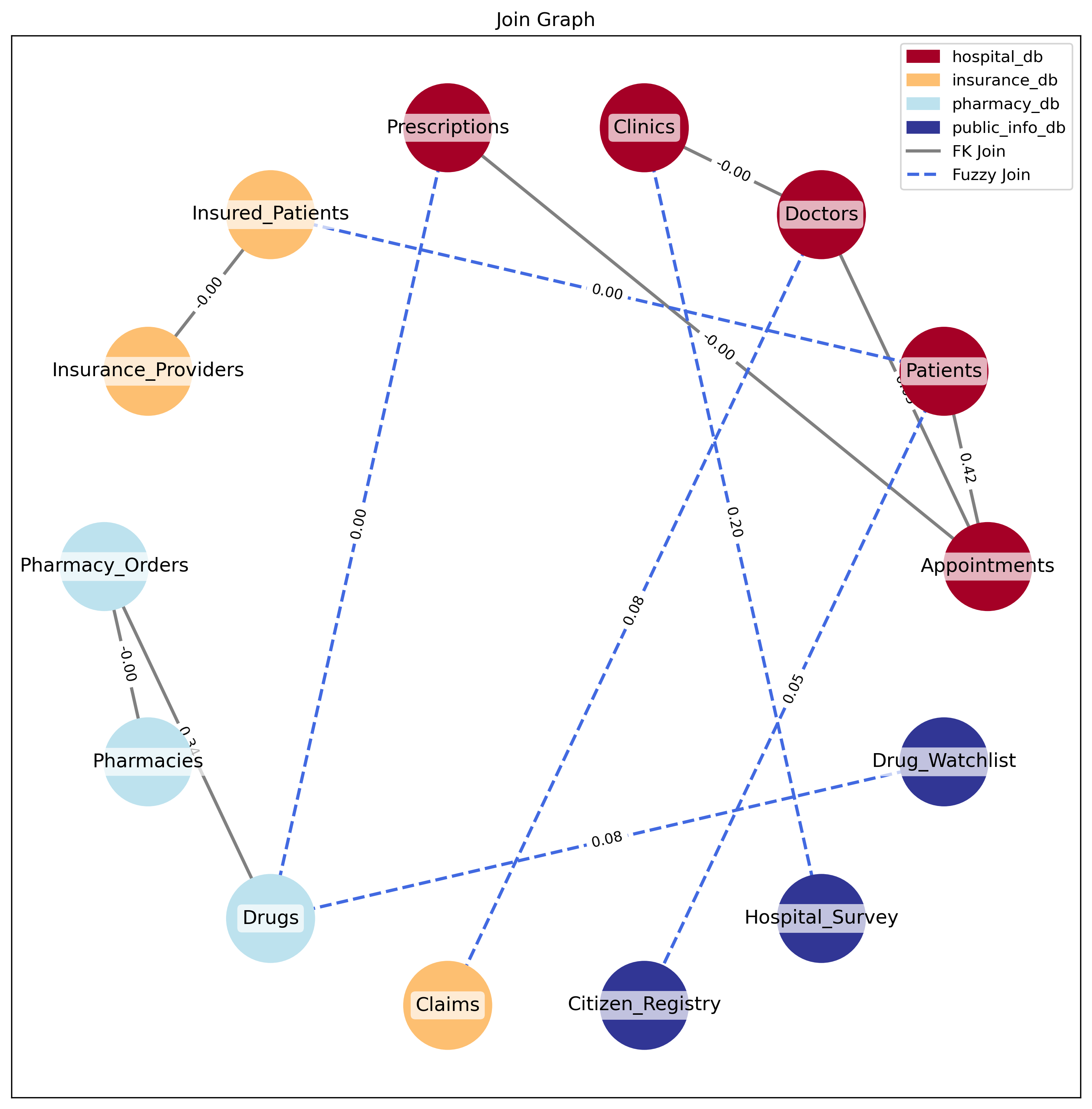}
    \caption{\textmd{Join graph constructed across all databases. Nodes represent tables, colored by database. Edges represent joinable column pairs, with solid gray lines for foreign key joins and dashed blue lines for fuzzy joins.}}
    \label{fig:join-graph}
\end{figure}

\subsection*{Example Scenario}

In this scenario, the user wants to determine the government evaluation score (from \texttt{Hospital\_Survey} in \texttt{public\_info\_db}) of the clinic that each doctor (from \texttt{Doctors} in \texttt{hospital\_db}) is affiliated with.

\paragraph{Join Path Selection}

It first performs a foreign key join between the \texttt{Doctors} and \texttt{Clinics} tables using the \texttt{clinic\_id} field. Then, it performs a fuzzy join between the \texttt{clinic\_name} and \texttt{hospital\_name} columns, with an overlap score of 0.81. The output includes each doctor’s name, their affiliated clinic, the clinic’s satisfaction score, and the fuzzy match score for the final join step.

\paragraph{Join Output}
Since current RDBMS doesn't support fuzzy record level join directly in databases, we output the joined results as a \texttt{Pandas Dataframe}. This is a function we are still working to refine. We aim to make the dataframe manipulation more stable for joins with many steps, and also to reduce the memory consumption during row manipulation.
\begin{table}[ht]
\centering
\renewcommand{\arraystretch}{1.3}  
\scriptsize
\begin{tabular}{|p{0.6cm}|p{0.6cm}|p{1.5cm}|p{1.8cm}|p{1cm}|p{0.7cm}|}
\hline
\textbf{Clinic ID} & \textbf{Doctor ID} & \textbf{Doctor Name} & \textbf{Clinic Name} & \textbf{Satisfaction Score} & \textbf{Fuzzy Score} \\
\hline
e2f9... & dd06... & Valerie Williams & Rodriguez--Johnson & 4.96 & 1.000 \\
2ea5... & 3a60... & Thomas Fleming & Fox--Medina & 4.48 & 1.000 \\
8762... & 1b8e... & Patrick Sheppard & Floyd--Hunt & 4.12 & 1.000 \\
a45f... & 47d7... & Alan Mercer & Reed, Blair and Allen & 4.66 & 0.810 \\
a45f... & e8e1... & Stephanie Jones & Reed, Blair and Allen & 4.66 & 0.810 \\
\hline
\end{tabular}
\vspace*{0.5em}
\caption{\textmd{Join result showing each doctor’s clinic and the corresponding government satisfaction score, along with the fuzzy match score. Column names are reformatted for improved readability.}}
\label{tab:scenario1-output}
\end{table}
\vspace*{-2em}

\section{Expected Contributions}
\subsection{Future Work: Many-to-Many Intermediate Joins on a Fuzzy Basis}
While our current system effectively identifies indirect join paths through intermediate tables, it is currently limited to one-to-one mappings. In future work, we plan to extend the system to handle many-to-many intermediate joins, which are common in real-world data scenarios such as mapping a set of columns (e.g., \texttt{ZIP}, \texttt{City}, \texttt{Address}) to another column (e.g., \texttt{Full\_Address}). This will require aggregations across multiple branches.

\subsection{Contributions}

This research contributes toward making relational databases more flexible and realistic in the face of schema inconsistencies and real-world data noise. By enabling automatic fuzzy joins across multiple tables—even without explicit foreign key constraints—we help lower the barrier for:
\begin{itemize}
    \item \textbf{Cross-domain data integration} in fragmented systems (e.g., healthcare, finance, government).
    \item \textbf{Exploratory data analysis} without prior knowledge of schema structure.
\end{itemize}
Our approach complements existing entity matching systems by extending join reasoning to the multi-table level, enhancing automation in join path discovery.

\section{Conclusions}
We present a fuzzy join discovery framework that combines column-level schema matching with row-level similarity scoring to infer joinable relationships across multiple tables and databases. By modeling the problem as a weighted join graph traversal, we allow for the discovery of indirect join paths through intermediate tables.

Our experiments on synthetic yet realistic databases demonstrate the system’s ability to recover meaningful joins and trace useful relationships in noisy environments. In the future, we aim to experiment the system on larger real-world databases, and expand the system to support many-to-many joins.

\bibliographystyle{ACM-Reference-Format}
\bibliography{references.bib}


\begin{thebibliography}{4}


\ifx \showCODEN    \undefined \def \showCODEN     #1{\unskip}     \fi
\ifx \showDOI      \undefined \def \showDOI       #1{#1}\fi
\ifx \showISBNx    \undefined \def \showISBNx     #1{\unskip}     \fi
\ifx \showISBNxiii \undefined \def \showISBNxiii  #1{\unskip}     \fi
\ifx \showISSN     \undefined \def \showISSN      #1{\unskip}     \fi
\ifx \showLCCN     \undefined \def \showLCCN      #1{\unskip}     \fi
\ifx \shownote     \undefined \def \shownote      #1{#1}          \fi
\ifx \showarticletitle \undefined \def \showarticletitle #1{#1}   \fi
\ifx \showURL      \undefined \def \showURL       {\relax}        \fi
\providecommand\bibfield[2]{#2}
\providecommand\bibinfo[2]{#2}
\providecommand\natexlab[1]{#1}
\providecommand\showeprint[2][]{arXiv:#2}

\bibitem[Bachmann(2023)]%
        {rapidfuzz}
\bibfield{author}{\bibinfo{person}{Max Bachmann}.} \bibinfo{year}{2023}\natexlab{}.
\newblock \bibinfo{title}{RapidFuzz: Fast fuzzy string matching in Python}.
\newblock \bibinfo{howpublished}{\url{https://github.com/rapidfuzz/RapidFuzz}}.
\newblock
\newblock
\shownote{Version 3.4.0, Accessed: 2025-06-11}.


\bibitem[Li et~al\mbox{.}(2021)]%
        {li2021auto}
\bibfield{author}{\bibinfo{person}{Peng Li}, \bibinfo{person}{Xiang Cheng}, \bibinfo{person}{Xu Chu}, \bibinfo{person}{Yeye He}, {and} \bibinfo{person}{Surajit Chaudhuri}.} \bibinfo{year}{2021}\natexlab{}.
\newblock \showarticletitle{Auto-FuzzyJoin: Auto-Program Fuzzy Similarity Joins Without Labeled Examples}. In \bibinfo{booktitle}{\emph{Proceedings of the 2021 International Conference on Management of Data (SIGMOD)}}.
\newblock


\bibitem[Peeters et~al\mbox{.}(2025)]%
        {peeters2025entity}
\bibfield{author}{\bibinfo{person}{Ralph Peeters}, \bibinfo{person}{Aaron Steiner}, {and} \bibinfo{person}{Christian Bizer}.} \bibinfo{year}{2025}\natexlab{}.
\newblock \showarticletitle{Entity Matching using Large Language Models}. In \bibinfo{booktitle}{\emph{Proceedings of the 2025 International Conference on Extending Database Technology (EDBT)}}.
\newblock


\bibitem[Reimers and Gurevych(2021)]%
        {reimers-2021-minilm}
\bibfield{author}{\bibinfo{person}{Nils Reimers} {and} \bibinfo{person}{Iryna Gurevych}.} \bibinfo{year}{2021}\natexlab{}.
\newblock \bibinfo{title}{Sentence-BERT: all-MiniLM-L6-v2}.
\newblock \bibinfo{howpublished}{\url{https://huggingface.co/sentence-transformers/all-MiniLM-L6-v2}}.
\newblock
\newblock
\shownote{Accessed: 2025-06-11}.


\end{thebibliography}

\end{document}